\documentclass[a4paper]{jpconf}
\usepackage{graphicx}
\usepackage{amsmath,amssymb}
\usepackage{dsfont}
\usepackage{slashed}
\usepackage{verbatim}
\usepackage{wrapfig}
\usepackage[    bookmarks,
                bookmarksopen = true,
               bookmarksnumbered = true,
                 linktocpage,
                 colorlinks = true,
                 linkcolor = blue,
                 urlcolor  = blue,
                 citecolor = blue,
                 anchorcolor = green,
                 hyperindex = true,
                 hyperfigures]
		{hyperref}

\newcommand{\D}{\mathcal{D}} 

\renewcommand{\d}{\textmd{d}}
\newcommand{\be}{\begin{equation}}
\newcommand{\ee}{\end{equation}}

\newcommand{\Z}{\mathcal{Z}}

\newcommand{\expv}[1]{\left \langle #1 \right \rangle}

\newcommand{\Regensburg}{Institute for Theoretical Physics,
Universit\"at Regensburg, D-93040 Regensburg, Germany.}

\begin{document}

\title{QCD phase diagram: overview of recent lattice results
}

\author{Gergely~Endr\H{o}di}

\address{\Regensburg}
        
\ead{gergely.endrodi@physik.uni-regensburg.de}

\begin{abstract}
Two parameters that have a strong influence on the finite temperature QCD transition, 
and play an important role in various physical scenarios are the quark density and 
the external magnetic field. The effect of these parameters on the thermal properties 
of QCD is discussed, and an overview of the latest lattice results is given.
\end{abstract}

\section{Introduction}

\begin{wrapfigure}{r}{7.6cm}
 \flushright
 \vspace*{-1.1cm}
 \includegraphics[width=7cm]{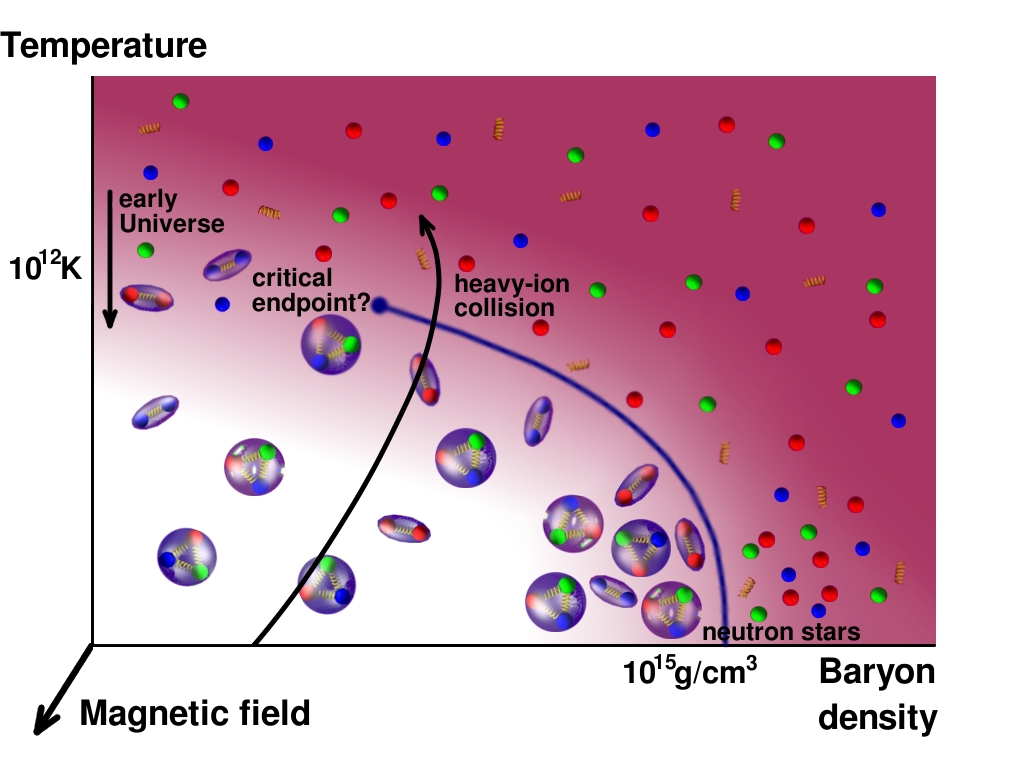}
 \vspace*{-.2cm}
 \caption{ \label{fig:pd}
  Conjectured QCD phase diagram.
 }
\end{wrapfigure}

Quantum Chromodynamics (QCD) is the theory of the strong interactions. Its elementary particles -- 
quarks and gluons -- cannot be observed directly at low energies, as they are confined to be 
constituents of bound states (hadrons). 
However, QCD is also an asymptotically free theory, which implies that at high energies 
quark and gluons are deconfined, and through a phase transition the quark-gluon plasma (QGP) 
phase is formed.
This transition between the hadronic and QGP phases can be represented on the phase diagram 
of QCD. A conjecture of how it might look is depicted in Fig.~\ref{fig:pd}.

To trigger deconfinement -- and, thus, to probe the high-energy regime of QCD -- an extreme 
environment is necessary, realized by high temperatures $T$ or baryon densities $\rho$. The 
characteristic values where the transition happens correspond to $T_c\approx 10^{12}\textmd{ K}$ and 
$\rho_c\approx 10^{15} \textmd{ g}/\textmd{cm}^3$. 
Temperatures of this strength were present in the early phase of the evolution of the universe, 
at about $1\mu s$ after the Big Bang. On the other hand, densities exceeding the above critical value are also 
conjectured to be present in the interior of compact, dense stellar objects, neutron stars. 
In both cases, strongly interacting matter enters its quark-gluon plasma phase and, thus, the 
properties of the transition between the hadronic and QGP phases are essential for the 
description of these systems. 
Furthermore, the quark-gluon plasma is also routinely reproduced in contemporary heavy-ion collisions. 
One relevant parameter here is the center-of-mass energy $\sqrt{s}$, which 
controls the temperature and the density of the created fireball. 
The Relativistic Heavy Ion Collider (RHIC) and the 
Large Hadron Collider (LHC) operate at high energies, where the initial excess of quarks over 
antiquarks is negligible compared to the total number of created particles, resulting at low 
net baryon densities. On the other hand, various experiments at lower energies aim to study the 
system at large net densities, for example RHIC II or the 
future Facility for Antiproton and Ion Research (FAIR). 
One of the most important objectives in the latter experiments is to identify a possible 
critical endpoint on the phase diagram, beyond which the transition becomes first order. 
Developing our theoretical understanding of the QCD phase diagram can prove highly beneficial for 
designing these next generation experiments.

Besides the temperature and the baryon density, there are additional parameters, which may be 
relevant for the above mentioned systems. Such a parameter is a background magnetic field $B$, which 
represents an important concept for early cosmology~\cite{Grasso:2000wj}, 
for a certain class of neutron stars called magnetars~\cite{Duncan:1992hi} 
and for non-central heavy-ion collisions~\cite{Kharzeev:2012ph}. 
Model descriptions of such collisions show that the induced magnetic field for typical events 
at RHIC or at the LHC may be of the order of several $m_\pi^2$. A magnetic field of this strength competes 
with the strong interactions and induces several new phenomena, for example it affects 
chiral symmetry and its restoration and (de)confinement, it introduces a spatial asymmetry and, thus, 
induces Lorentz-symmetry breaking expectation values and it also changes the hadron spectrum 
considerably. For recent reviews on the subject, see, e.g., Refs.~\cite{Kharzeev:2012ph,Bali:2013cf}. 
For the phase diagram, the magnetic field introduces a third axis and defines 
the three-dimensional phase diagram of QCD, see Fig.~\ref{fig:pd}.

Our current knowledge about the phase diagram is still very limited. In fact, most of the 
information illustrated in Fig.~\ref{fig:pd} is based on descriptions of strongly interacting matter 
in terms of low energy hadronic models (low temperatures, densities or magnetic fields) 
or perturbation theory (high temperatures, densities or magnetic fields). 
However, in order to study the transition region, where the hadronic description is no longer valid, 
but perturbation theory is not yet feasible due to the large QCD coupling, one needs to take 
into account the non-perturbative nature of QCD. 
The best non-perturbative and systematically adjustable approach to QCD is lattice field theory, 
where the field degrees of freedom are discretized on a four-dimensional space-time grid. 
Through the lattice approach we can study the strong dynamics solely from first principles -- 
i.e., from the QCD Lagrangian.

The primary quantity in this approach is the QCD partition function 
(in the grand canonical formalism, where the density is controlled by a chemical potential $\mu$), 
which is given by the functional integral over the gluon fields $U$,
\be
\Z = \int \D U e^{-S_g} \det\left[ \slashed{D}(B,\mu) + m \right],
\label{eq:Z}
\ee
where $S_g$ is the gluonic action and $\slashed{D}$ the Dirac operator. 
The lattice regularization defines $\Z$ through the limiting procedure of considering finer and finer grids.
This amounts to increasing the number $N_s$ of lattice points to infinity and taking the lattice spacing 
$a$ to zero in a manner that the physical size $N_sa$ is fixed. 
This is called the continuum limit, which requires simulations of bigger and bigger lattices, 
increasing the computational costs drastically. 
Another, computationally demanding aspect is the tuning of the bare quark masses $m$ to their 
physical values. This procedure is performed such that the lattice measurements of physical observables -- 
e.g. hadron masses at zero temperature -- coincide with their experimental values. Tuning $m$ in this 
manner makes the condition number of the matrix $\slashed{D}+m$ large and, accordingly, 
the simulation very expensive. 
Nevertheless, performing the continuum limit and using physical quark masses in the simulation are 
necessary ingredients to 
produce reliable precision results that can be compared to experiments or models. 

In this talk, I will present recent lattice results about the phase diagram, concentrating first 
on nonzero chemical potentials, then turning to the case of nonzero magnetic fields.

\section{Phase diagram at zero and low chemical potentials}

The most important properties of the transition separating the hadronic and QGP phases are 
the characteristic temperature $T_c$ and the order of the transition. 
To determine these, an observable sensitive to the transition is necessary. 
Such observables are the chiral condensate $\bar\psi\psi=\partial \log\Z/\partial m$ and the chiral 
susceptibility $\chi = \partial^2 \log\Z / \partial m^2$. 
The latter is 
similar to a specific heat and exhibits a peak-like structure in the transition region. 
The transition temperature is determined as the position of the peak, 
whereas the order of the transition can be extracted via a finite size scaling analysis of the height $h$ 
of the peak: $h\propto V^\alpha$, where $\alpha=1$ indicates a first-order, $0<\alpha<1$ a second-order 
phase transition, and $\alpha=0$ an analytic crossover, which involves no singularity.
Such a finite size scaling analysis was performed in Ref.~\cite{Aoki:2006we} revealing the transition 
at $\mu=B=0$ to be an analytic crossover. One implication of this is that $T_c$ is not unique but 
can depend on the observable used for its definition. 
The transition temperature defined via various observables was determined using 
different discretizations of the fermionic action~\cite{Aoki:2006br,Aoki:2009sc,Borsanyi:2010bp}
\cite{Bazavov:2011nk}, giving values scattered around $T_c\approx 160 \textmd{ MeV}$. 

\begin{figure}[ht!]
 \centering
 \vspace*{-1.5cm}
\begin{tabular}{p{0.5\textwidth} p{0.5\textwidth}}
  \vspace{.8cm} \includegraphics[height=5cm]{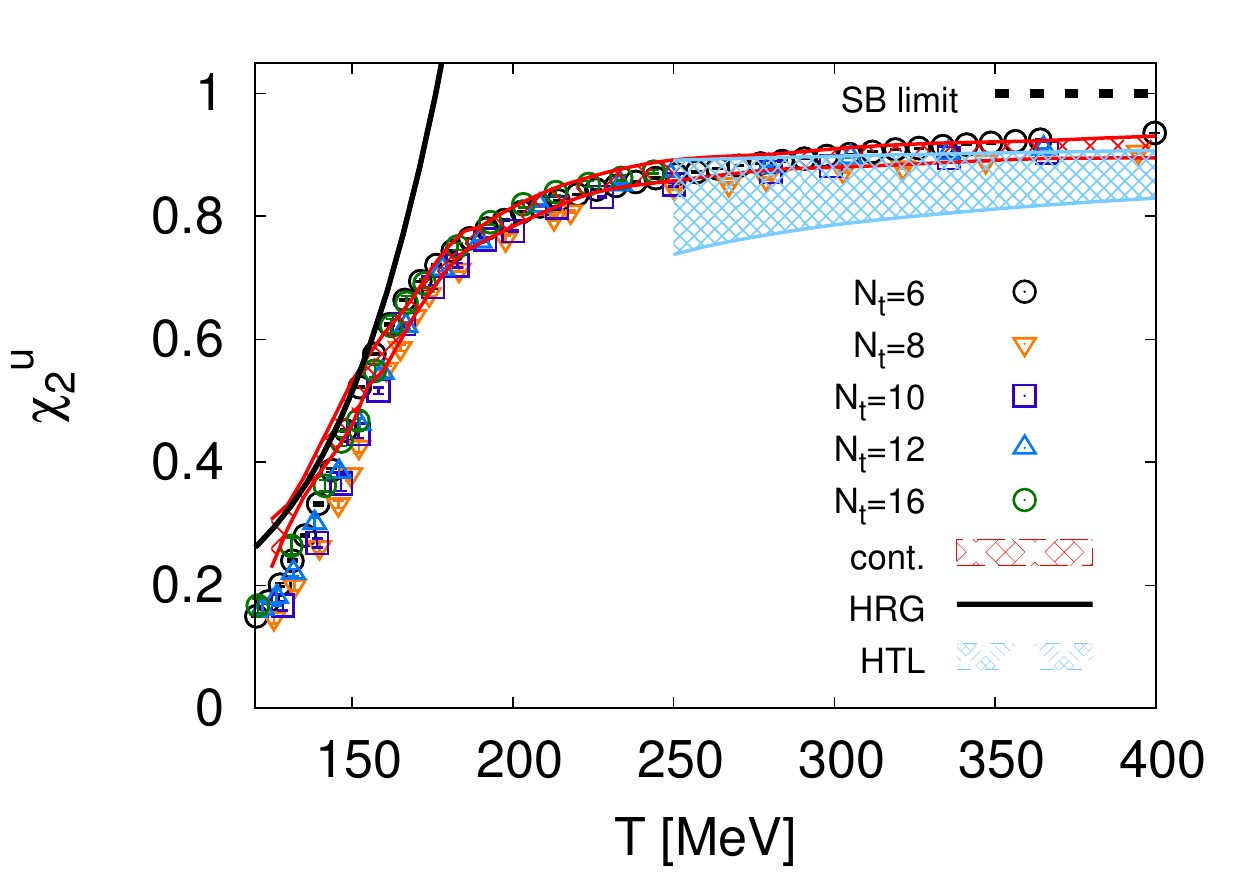} &
  \vspace{0pt} \includegraphics[height=6.3cm]{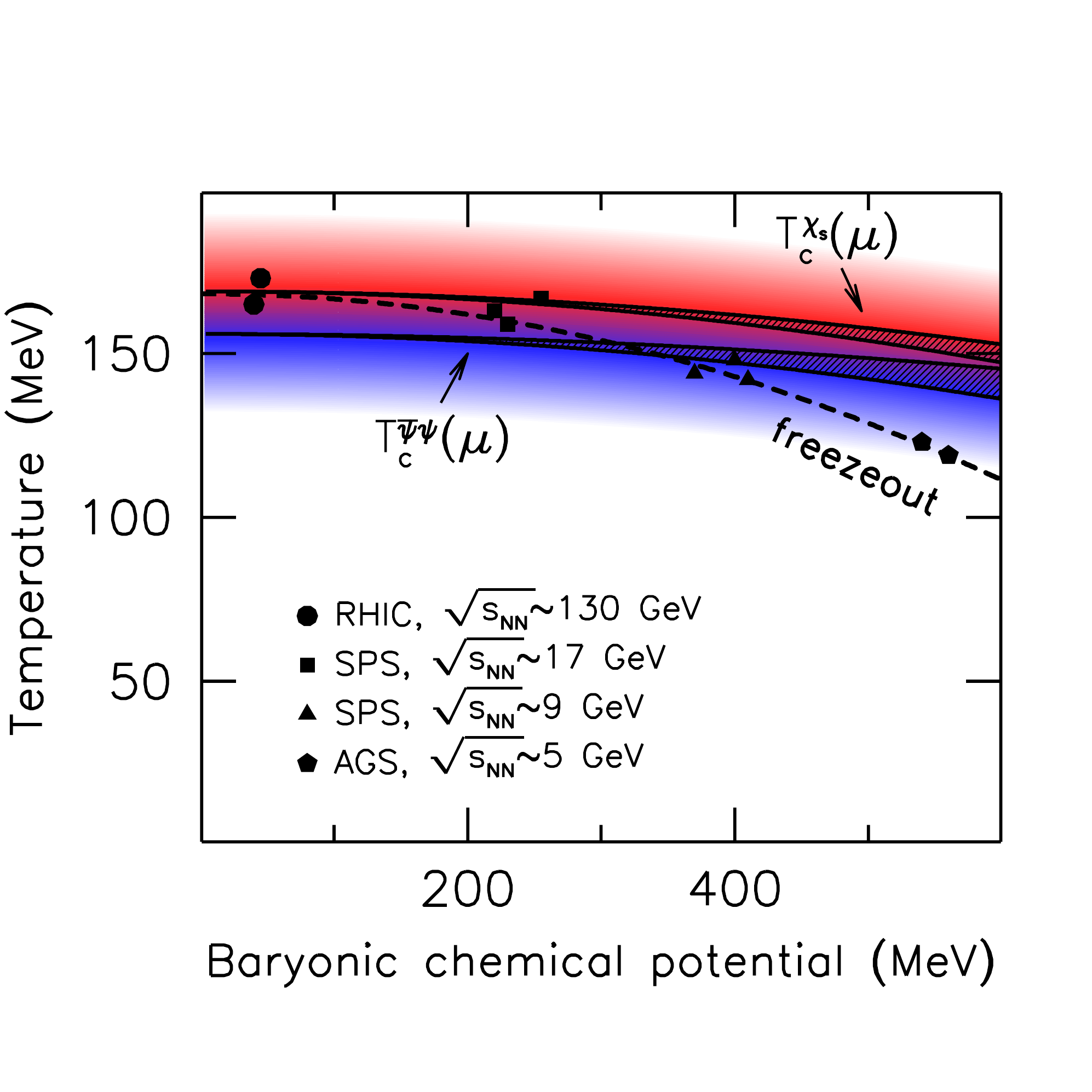}
\end{tabular}
 \vspace*{-.8cm}
 \caption{ \label{fig:mu2}
  Left panel: continuum limit of the up quark number susceptibility compared to a hadronic model (solid line) and perturbation theory (blue dashed region). Right panel: QCD phase diagram for small chemical potentials, 
with two definitions of $T_c$. 
 }
\end{figure}

The determination of these characteristics at $\mu>0$ is, unfortunately, much more difficult. 
The lattice approach relies on the importance sampling of the partition function of Eq.~(\ref{eq:Z}) based on the 
weight $e^{-S_g} \det [ \slashed{D}+m ]$. This weight, however, becomes complex for any $\mu>0$, rendering 
standard Monte-Carlo methods useless. Several methods have been developed to alleviate this complex action 
problem (see, e.g., Refs.~\cite{deForcrand:2010ys,Aarts:2013bla}), 
still, they are only applicable at low $\mu$.
One possibility is to perform 
a Taylor-expansion in the chemical potential at $\mu=0$. The leading-order dependence on the chemical potential 
is encoded in the quark number susceptibilities $\chi^f_2 = \partial^2 \log\Z / \partial \mu_f^2$, 
which are closely related to fluctuations of conserved charges and, as such, 
can be used for a comparison with heavy-ion collisions ($f$=$u$,$d$,$s$ labels quark flavors). 
The left panel of Fig.~\ref{fig:mu2} 
shows the continuum limit of $\chi_2^u$ in the transition region, as obtained in 
Ref.~\cite{Borsanyi:2011sw}. 
The dependence of $T_c$ on $\mu$ can also be extracted via correlators of certain observables 
with $\chi_2^u$. The expansion in the chemical potential reads 
$T_c(\mu) = T_c(0) \cdot [1-\kappa\cdot \mu^2/T_c^2(0) ]$.
Due to the crossover nature of the transition, the coefficient $\kappa$ also depends 
on the observable used for its definition. Using $\bar\psi\psi$ and $\chi_2^s$ we obtained 
the values~\cite{Endrodi:2011gv} corresponding to the phase diagram shown in the right panel of 
Fig.~\ref{fig:mu2}. The results are in agreement with Ref.~\cite{Kaczmarek:2011zz}, 
where a different fermionic discretization was used. 
Note that equation of state-related observables (pressure, energy density, etc.)~\cite{Borsanyi:2010cj} 
tend to give somewhat larger values for the curvature~\cite{Borsanyi:2012cr}.

In Ref.~\cite{Endrodi:2011gv} we also determined the strength of the transition to leading order, implying 
a mild weakening of the transition in the low-$\mu$ region. This suggests to exclude the possibility of 
a critical endpoint at these low values of the chemical potential. 

\section{Phase diagram at nonzero magnetic fields}

Next, I turn to present recent results about the phase diagram at $\mu=0$ and $B>0$. Here
we consider a constant external magnetic field, i.e. 
the dynamics of $B$ is neglected. In the lattice simulation one needs to take into account 
the $B$-dependence of the fermion determinant in Eq.~(\ref{eq:Z}). This way the magnetic field 
couples directly to quarks, but it also changes the gauge backgrounds, revealing its indirect coupling 
to gluons. It is important to stress here that -- contrary to the case of nonzero chemical potentials 
-- the magnetic field does not lead to a complex action problem, therefore, 
the full $T-B$ phase diagram is accessible for lattice simulations. 

\begin{figure}[ht!]
 \centering
 \vspace*{-.2cm}
 \mbox{
 \includegraphics[height=5.cm]{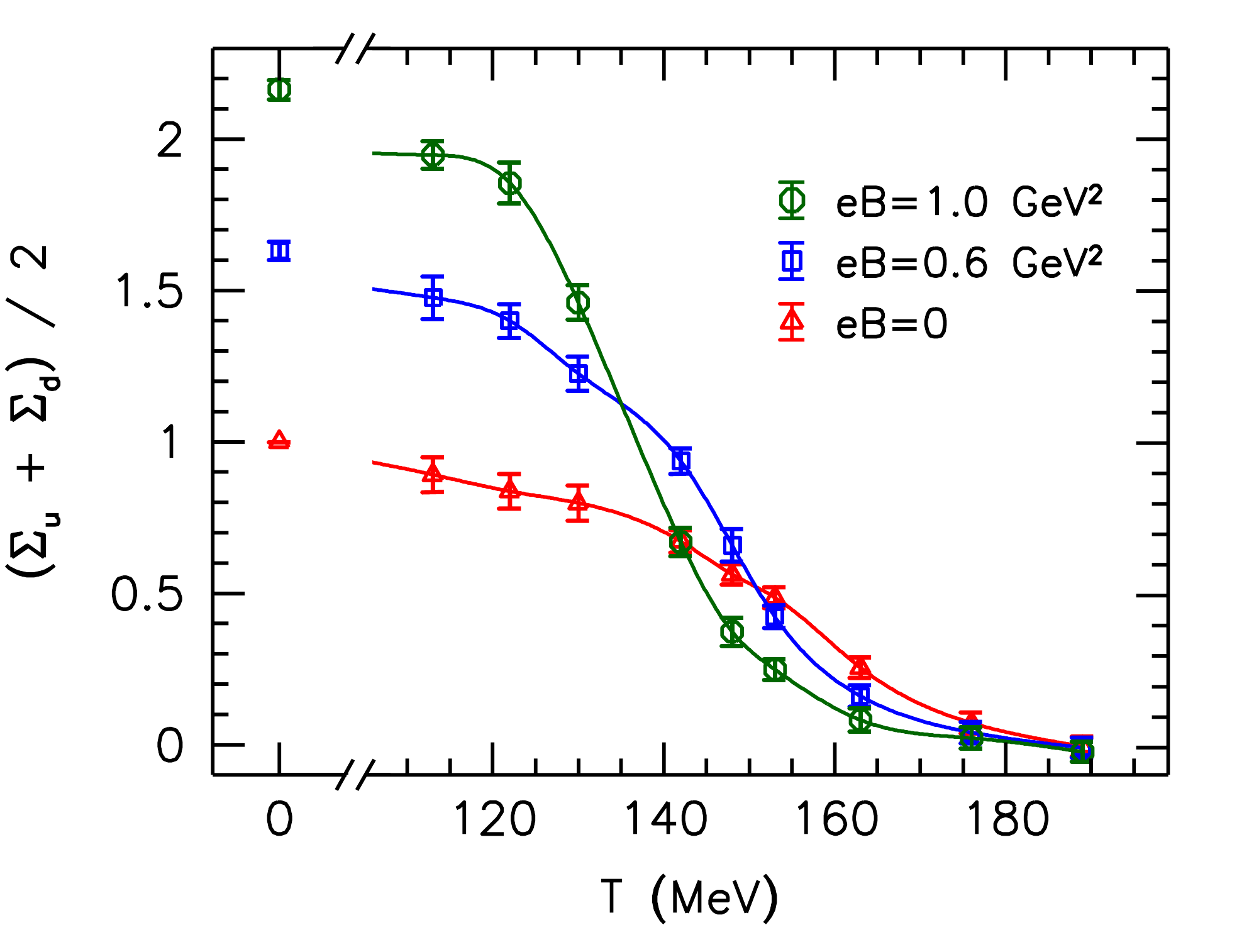} \quad \quad
 \includegraphics[height=5.15cm]{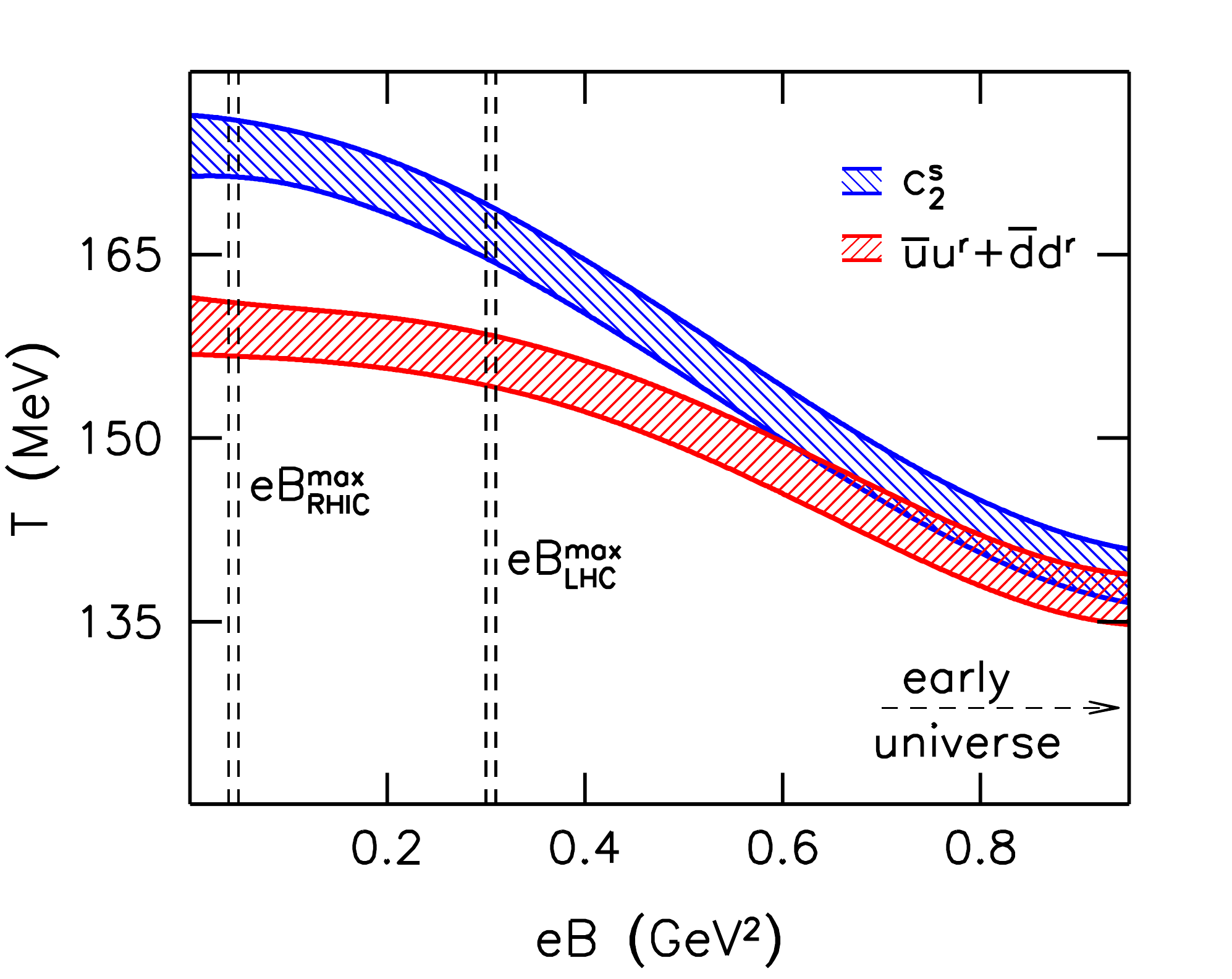}
 }
 \vspace*{-.3cm}
 \caption{ \label{fig:imc}
  Left panel: catalysis of $\bar\psi\psi$ at low temperatures and inverse catalysis around $T_c$. 
  Right panel: phase diagram in the $T-B$ plane, with $T_c$ defined as the inflection point of the condensate 
(red curve) or that of the strange quark number susceptibility (blue curve).
 }
\end{figure}

A pronounced effect of the magnetic field on QCD dynamics is the {\it magnetic catalysis} of the 
quark condensate, which implies $\bar\psi\psi$ to 
increase with growing $B$ at zero temperature. 
Magnetic catalysis at $T=0$ is a well-established phenomenon; it was observed in various effective theories 
and models, and confirmed by lattice simulations (see the recent review~\cite{Shovkovy:2012zn}). 
On the other hand, contrary to most of the low-energy models, 
our continuum extrapolated lattice simulations at physical quark masses 
have also shown that $\bar\psi\psi$ exhibits a non-monotonic dependence on the magnetic field around $T_c$, with a 
certain region, where it actually decreases with growing $B$~\cite{Bali:2011qj}. 
This behavior -- which we named {\it inverse magnetic catalysis} -- operates in a way that 
the inflection point of the condensate (which we use to define $T_c$ here) is shifted towards smaller 
temperatures (see left panel of Fig.~\ref{fig:imc}). As a result, 
the transition temperature $T_c(B)$ is reduced as the magnetic field increases, see right panel of 
Fig.~\ref{fig:imc}, where the phase diagram is shown.

Both the inverse catalysis around $T_c$ and the reduction of $T_c(B)$ were
rather unexpected, since 
most low-energy effective theories and models predicted the magnetic catalysis to prevail for all 
temperatures, and, $T_c(B)$ to increase. 
This prompted us to study the mechanism behind inverse catalysis in more detail. In 
Refs.~\cite{Bali:2011qj,Bali:2012zg} we found that this behavior is peculiar for light quarks, and it disappears 
for heavier-than-physical quark masses (see also Ref.~\cite{D'Elia:2010nq}). Moreover, we also found the inverse catalysis to be related 
to the indirect coupling of the magnetic field to gluons. 
There is a straightforward way to distinguish between the quark (direct) and gluonic (indirect) contributions 
to the condensate in terms of the partition function, Eq.~(\ref{eq:Z})~\cite{D'Elia:2011zu}. 
The expectation value of $\bar\psi\psi$ 
reads
\be
\expv{\bar\psi\psi} = \frac{1}{\Z} \int \D U e^{-\beta S_g} \, \underbrace{\det [ \slashed{D}(B)+m ]}_{\rm indirect} \,\underbrace{\Tr [\slashed{D}(B)+m]^{-1}}_{\rm direct}.
\ee
The direct part is the dependence of the operator on $B$, whereas the indirect part is 
the dependence through the determinant. We found that the direct term always strives 
to increase the
\begin{wrapfigure}{l}{6.8cm}
 \centering
 \vspace*{-.2cm}
 \includegraphics[height=5.15cm]{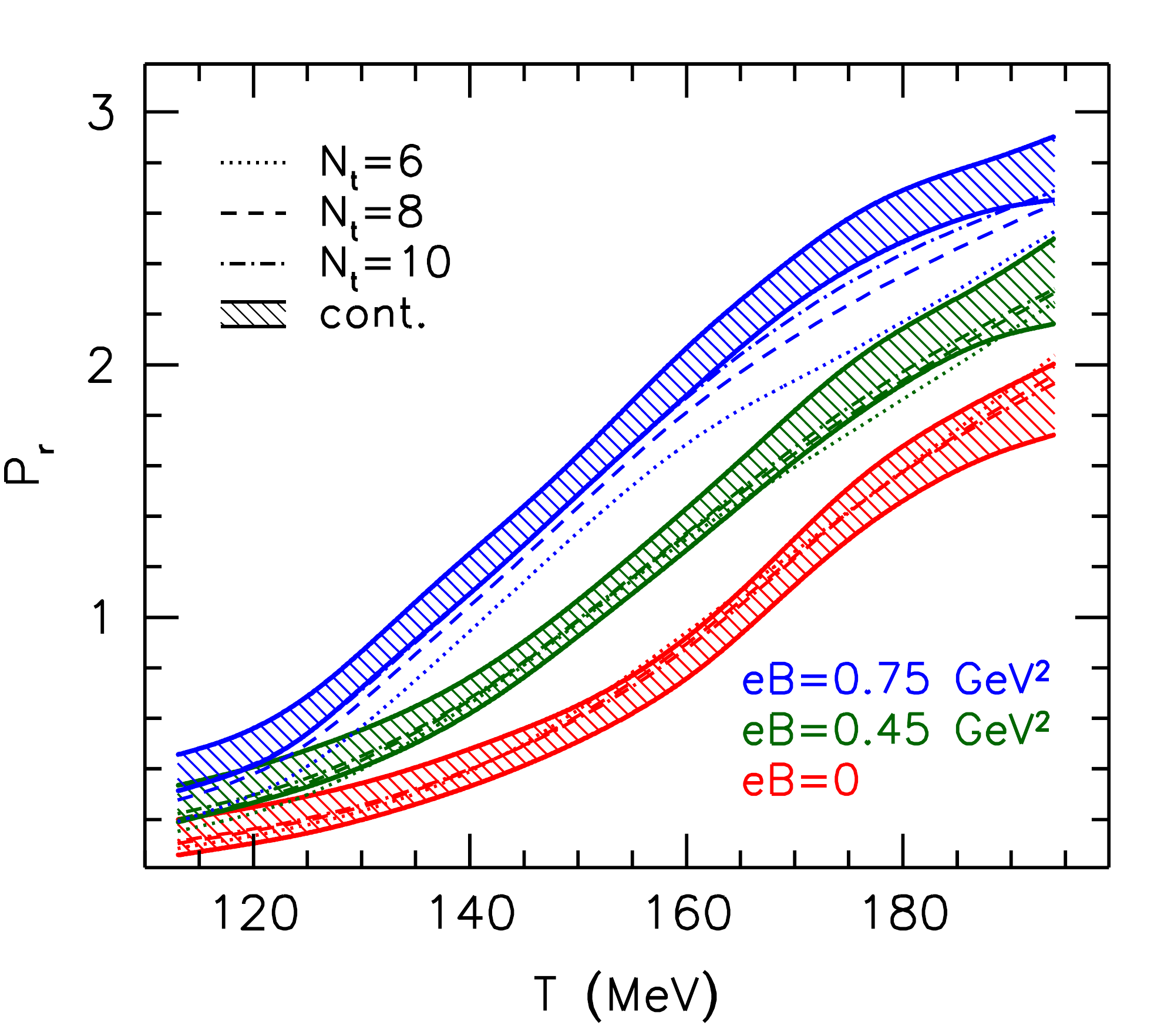}
 \vspace*{-.4cm}
 \caption{ \label{fig:ploop}
Continuum limit of $P$ as a function of $T$ for various magnetic fields.
 }
\vspace*{-.3cm}
\end{wrapfigure}
expectation value, whereas 
the indirect contribution tends to decrease it around $T_c$, if the quark mass $m$ is 
small~\cite{Bruckmann:2013oba}. 
It is then natural to look for the relevant degrees of freedom in the gauge background that this 
indirect effect couples to. The most important gauge degree of freedom around $T_c$ 
is the Polyakov loop, the parallel transport winding around the temporal direction of the lattice 
$P=\Tr \exp \big[\int_0^{1/T}\! A_4 \,\d t\big]$. Therefore, we proceeded by determining it in the continuum limit for various 
magnetic fields, see Fig.~\ref{fig:ploop}. The results show that $P$ depends predominantly on the magnetic 
field around $T_c$, namely it is increased by $B$. At the same time, the inflection point of $P$ is shifted to 
lower temperatures, revealing that around $T_c$, the magnetic field favors gauge 
configurations corresponding to the deconfined phase, i.e. with small condensate. This leads to 
the inverse catalysis of $\expv{\bar\psi\psi}$ in the transition region.

\section{Summary}

In this talk I discussed the effect of chemical potentials and background magnetic fields on 
QCD thermodynamics. The former introduces a complex action problem and prohibits direct simulation 
of the theory. To circumvent this problem we applied a Taylor-expansion in $\mu$. 
Our continuum extrapolated results at physical quark masses reveal the phase diagram 
for low chemical potentials to be as depicted in the right panel of Fig.~\ref{fig:mu2}. 
On the other hand, for $\mu=0$ and $B>0$, the weights remain real, allowing for the determination of the 
whole phase diagram by conventional lattice techniques.
A pronounced effect here is the inverse magnetic catalysis around $T_c$, and the reduction of 
the transition temperature with growing $B$, see Fig.~\ref{fig:imc}. \\

\noindent{\bf Acknowledgments. } I would like to thank the FAIRNESS 2013 organizers for the inspiring meeting. 
I am grateful to my colleagues with whom I collaborated on various projects 
presented in this talk.
My work was supported by the EU (ITN STRONGnet 238353). \\

\bibliographystyle{iopart-num}
\bibliography{fairness13_iop}

\end{document}